\documentclass[12pt]{article}
\usepackage{graphicx}
\begin{document}

{\bf Opinion Dynamics and Sociophysics}

\bigskip

D. Stauffer

\bigskip
Institute for Theoretical Physics, Cologne University, 

D-50923 K\"oln, Euroland

\bigskip
{\bf \large Article Outline}
\bigskip

Glossary

I. Definition and Introduction

II. Schelling Model

III. Opinion Dynamics

IV. Languages, Hierarchies and Football

V. Future Directions

\bigskip
{\bf \large Glossary}
\bigskip

{\bf Cluster}

Clusters are sets of neighbouring sites of the same type.

\medskip
{\bf Ising model} 

Each site carries a magnetic dipole which points up or down; neighbouring
dipoles ``want'' to be parallel.

\medskip
{\bf Opinion dynamics} 

How do people change opinions? Simulations usually ignore all details of the 
brain and represent the opinion by one or several numbers which can be changed 
due to contact with others.

\medskip
{\bf Schelling model}

People belonging do different groups may produce segretated neighbourhood just 
by their personal preferences, not by outside force.

\medskip
{\bf Sociophysics}

Application of methods from (mostly statistical) physics to human relations;
can be traced centuries backwards.

\bigskip
\section{Definition and Introduction}

\begin{figure}[hbt]
\begin{center}
\end{center}
\caption{Ising model after 20 Glauber kinetic steps per site on a $500 \times 
500$ square lattice at $k_BT/J = 2$. We start from a random distribution
of equally many black and white sites. Figs.1, 2, 3 omitted for arXiv; their 
small are not readable on some computers and show the standard phase separation
or clustering of Ising model..
}
\end{figure}

\begin{figure}[hbt]
\begin{center}
\end{center}
\caption{As Fig.1 but after 2000 instead of 20 iterations.}
\end{figure}

The application of concepts from the natural sciences to social sciences,
partly to be reviewed here, is at least 25 centuries old. Then the Greek
philosopher Empedokles stated (according to J. Mimkes) than humans are like
liquids: Some mix easily like wine and water, and others like oil and water
refuse to mix. We start with the Schelling model of 1971 which implemented
this idea, and its criticism. Then we will review opinion 
dynamics in large populations, summarising only shortly other aspects like 
self-organisation of hierarchies or competition between human languages.

Humans do not like to be treated like a number, and indeed the human brain is 
much more complex than a binary variable (called ``spin'' by physicists) which
is either +1 or --1. We do not deal here with the psychological processes
of an individual but with mass psychology, and this author learned half a
century ago in school that mass psychology is different from individual 
psychology: The law of large numbers averages out over individual fluctuations
and makes general trends more clearly visible. Thus what we call today 
statistical physics plays a useful rule, and social scientists 
\cite{schelling,jones} have applied it, without knowing then that they dealt 
with an Ising model of ferromagnets. 

The astronomer Halley, best known through 
his comet, tried to establish mortality tables already three centuries ago.
Of course, the time of death of one given individual is usually difficult to 
predict but averaged over millions of people the statistical offices of 
many countries prepare regularly life tables which tell us how probable it is 
for a newborn child to live up to $x$ years, provided there are no changes
of the mortalities in the coming decades. Insurance for automobiles is another
example: We do not want to produce accidents, but we know that they happen,
and take precautions for their financial consequences. Thus the whole 
insurance industry is based on treating humans like numbers, ignoring their 
individuality.

Finally, human opinions are often fluctuating and ill-defined, but nevertheless
in elections people cast one choice, out of a limited number of choices. And
election results belong to those social data for which we have lots of 
accurate numbers, based on large populations. 

Thus it is not at all the merit (or ignorance) of physicists which treats humans
like numbers; this method has a very long tradition and is an indispensable 
part of modern life.

\section{Schelling Model}

\subsection {Ising simulations}

Following (but not citing) Empedokles, the later economics Nobel laureate
Schelling
\cite{schelling} asked whether the racial segregation in American cities can
emerge from intrinsic behaviour of the individual people, instead of or in 
addition to extrinsic reasons like discrimination, rent differences, etc. In 
particular, can ``black'' ghettos in the predominantly ``white'' USA arise 
just because people prefer to have neighbours of their own group over 
neighbours from the other group? In many other countries we find many other
types of residential segregation, based on religion, ethnicity, .... In 
physics, such a process is easily simulated through the two-dimensional
Ising model, as shown in Figs.1 and 2.

In this Ising model, each site on a square lattice carries a variable $S_i = 
\pm 1$, and each pair $<i,k>$ of nearest neighbours produces an ``energy''
$-JS_iS_k$ with some proportionality constant $J$. The total energy $E$ 
(= total unhappiness) is the
sum of this pair energy over all neighbour pairs of the lattice. In physics,
different distributions of the ``spins'' $S_i$ are realised with a probability
proportional to exp($-E/k_BT)$ where $T$ is the absolute temperature and $k_B$
the Boltzmann constant. There is no need to worry about values for
$T, \; k_B, \; J$ since the only relevant quantity is the ratio $k_BT/J$, taken
as 2 in these pictures. The ``Glauber'' kinetics is simulated on the computer by
flipping a spin if and only if a random number between 0 and 1 is smaller than
the probability $\exp(-\Delta E/k_BT)/[1+\exp(-\Delta E/k_BT]$. The Fortran 
program  contains less than 40 lines and takes a few seconds:

{\small
\begin{verbatim}
      parameter(L=500,Lmax=(L+2)*L)
      dimension is(Lmax),iex(9)
      byte is
      data t,max,ibm/2.00,2000,1/
      print *, L,max,ibm,t
      Lp1=L+1
      L2pL=L*L+L
      do 1 i=1,Lmax
        is(i)=-1
        ibm=ibm*16807
 1      if(ibm.gt.0) is(i)=1
      do 2 ie=1,9
        ex=exp(-2*(ie-5)/t)
 2      iex(ie)=(2.0*ex/(1.0+ex) - 1.0)*2147483647
      ibm=2*ibm+1
      do 3 mc=1,max
        do 4 i=Lp1,L2pL
          ie=5+is(i)*(is(i-1)+is(i+1)+is(i-L)+is(i+L))
          ibm=ibm*16807
 4        if(ibm.lt.iex(ie)) is(i)=-is(i)
        mag=0
        do 6 i=Lp1,L2pL
 6        mag=mag+is(i)
 3      if(mc.eq.(mc/100000)*100000) print *, mc,mag
      do 5 i=Lp1,L2pL
        if(is(i).ne.1) goto 5
        iy = i/L
        ix=i-L*iy
        print *, ix, iy
 5    continue
      stop
      end
\end{verbatim}}

Such models and programs are taught in courses on computational or theoretical
physics all over the world; the model was published in 1925. If in the 
above flipping probability the denominator is omitted one gets the Metropolis
kinetics. If instead of flipping one spin we exchange two opposite spins,
we get the Kawasaki dynamics. For Glauber or Metropolis, after very long
times (measured by the number of sweeps through the lattice) one of the two 
possibilities dominates at the end, if $T$ is not larger that the critical
temperature $T_c$, with  $2J/k_BT_c = \ln(1 + \sqrt 2) \simeq 0.44 $ known 
since 1940. For Kawasaki dynamics the fraction of 
black sites remains constant, and we get two large domains. 
For higher temperatures above $T_c$ only small clusters and no large domains
are formed, Fig.3.

\begin{figure}[hbt]
\begin{center}
\end{center}
\caption{As Fig.2 but at $k_BT/J$ = 3 instead of 2. Only small clusters 
and no large domains are formed. After 200 and 20,000 iterations the 
pictures look similar to this one made after 2000 iterations. 
}
\end{figure}

In this Ising model, two neighbouring spins have due to their interaction
$-JS_iS_k$ a higher probability to belong to the same group than to belong to
the two different groups. If the difference between these two probabilities
is large enough, $T < T_c$, domain sizes can grow to infinity in an infinite 
lattice, Figs. 1 and 2, while only small clusters are formed in Fig.3 for 
smaller differences in the probabilities, $T > T_c$. That these probabilities, 
controlled through $-J/k_BT$, lead to these different regimes, separated by a 
sharp phase transition at $T = T_c$, is not obvious from the definition of the
interaction $J S_iS_k$, took physicists many years to find, and is typical of 
complex systems.

The social meaning of temperature $T$ is not what we hear in the weather 
reports but an overall approximation for all the more or less random events 
which influence our decisions but are not explicitely included in the model.
For residential segregation the model only counts how many neighbours of which
group one has. But not all people of one group are alike, housing in
different parts of a city costs different amounts of money, some parts are 
more beautiful then others, and job hunting may force us into a temporary 
residence of a new city which does not conform to our wishes. In this way,
a positive temperature allows for rare moves which increase the energy, i.e. 
we move to a new residence where the neighbourhood composition along makes
us less happy. At zero temperature, the Ising model does not properly order 
into one or two ``infinite'' domains.

\subsection{Schelling's version and later improvements}

Schelling \cite{schelling} avoided probabilistic rules and thus counted 
neighbours $S_k = \pm 1$ at zero temperature. Then it does not matter if all 
neighbours or only a majority of them belong to the own group. Thus people are 
defined as happy if at least half of the neighbours belong to the own group, 
and as unhappy otherwise (i.e. if the majority belong to the opposite 
group). Unhappy people move to the nearest place where they are happy. Since
Schelling moved only one person (or family) at a time, and made no exchange of 
two people simultaneously as in Kawasaki kinetics, he introduces a large 
fraction of empty residences. Thus at each step, one unhappy person or family 
moves into the closest vacancy where life would be happy. 

This model, and also many variants \cite{schelling,fossett}, fail to give 
large domains; only small clusters are seen. In reality, Harlem in Manhattan
(New York), is not a cluster of a few houses but extends over many square 
kilometers. Thus the original version does not give the desired results.
Large domain are formed if people also change residences if this brings
no improvement \cite{kirman} (hardly a realistic assumption) or at a finite 
temperature \cite{solomon}. The latter paper also gives some alternatives 
to the Schelling model which also allow for large domains, and a simple
example of s finite cluster where everybody is ``happy'' and which therefore
never grows or dissolves on its own ``will''. 
 
Much earlier and simpler is the zero-temperature version of Jones \cite{jones} who at
each iteration removes a random fraction of the people and fills the vacancies 
with people who are there happy in the Schelling sense. This randomness, just
like the finite temperature, leads to large domains as desired. Neither 
physicists nor social scientists have taken much note of \cite{jones}.
The history of the Schelling model is an example how the lack of communication
between disciplines has hampered progress in research, even very recently 
\cite{kirman,solomon}. Only computational statistical physicists know 
everything. (Jones in footnote 4 of \cite{jones} also mentions a 
probabilistic version closer to the Ising model.)
 
For finite temperatures, \cite{solomon} follows the above Glauber dynamics, but 
instead of an energy $E$ uses a variable which is 0 or 1 depending on the 
happiness of the residents. Moving from one place to the other then depends 
exponentially on the ratio of this variable to $k_BT$, instead of on the ratio 
$\Delta E/k_BT$. Many variants are possible, e.g. in the treatment of neutral 
cases \cite{kirman,mueller} where the number of neighbours of both groups is 
exactly the same.

But we are on safer grounds and can use decades of physics research if we use
the normal Ising model, or its generalisation to $Q$ different groups, the
$Q$-state Potts model. Then \cite{ortmanns,schulze} implemented a suggestion
of Weidlich \cite{weidlich} that people slowly learn to live together with
neighbours from the other group. Thus $T$ not only takes into account the
various accidents from outside the model, but also measures the tolerance:
The higher $T$ is the more are people willing to live in neighbourhoods of 
the other group. In the limit $T=\infty$ the neighbours would not matter at 
all, for intermediate $T$, Fig.3 showed small clusters but no large domains, 
and for low $T$ the domains grow to infinite sizes on an infinite lattice. 
The learning suggested by Weidlich thus means that this parameter $T$ (= 
temperature or tolerance) no longer is kept constant but slowly increases.

For an Ising model, \cite{ortmanns} showed how an initial large domain 
dissolves if the temperature is slowly increased from below to above $T_c$. 
More realistically, for five
(instead of only two) different groups in a modified 5-state Potts model, 
\cite{schulze} increased $T$ from low to high values and showed that with a 
slow increase one has appreciable domain formation during intermediate times, while with
a fast increase this segregation is mostly avoided, Fig.4.
   
\begin{figure}[hbt]
\begin{center}
\includegraphics[angle=-90,scale=0.5]{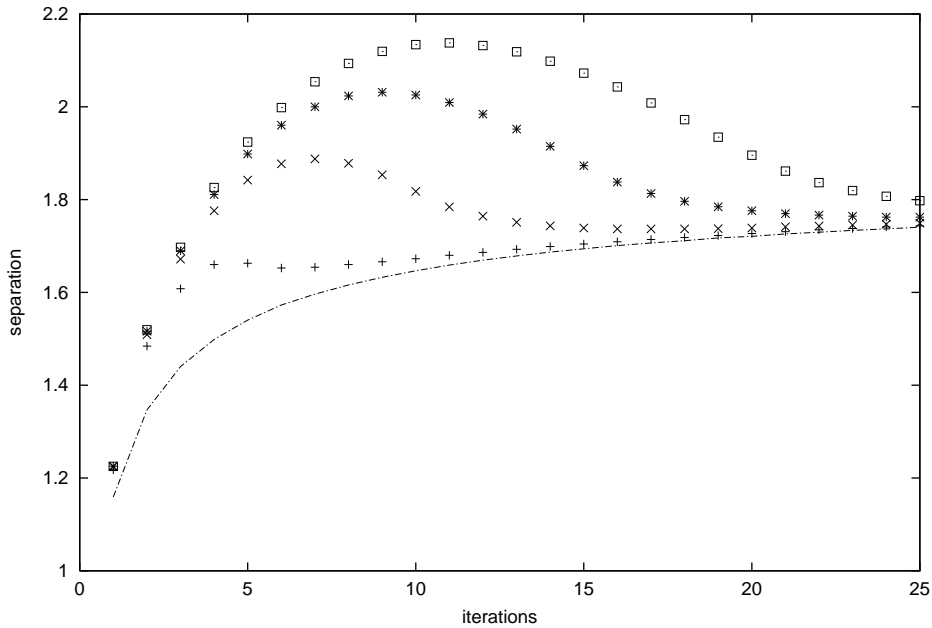}
\end{center}
\caption{Amount of neighbours of the same type in a Potts model of five groups,
normalised to unity for the initial random distribution. The temperature or 
tolerance increases from low to high values, slowly in the top curves, and 
fast in the lower curves; the latter mostly avoid the segregation into 
different group. (The lowest line holds for a constant high temperature.). 
From \cite{schulze}. 
}
\end{figure}

\section{Opinion Dynamics}

The following section describes several rules for simulated people to change 
opinions; each of these rules is applied again and again to these agents until
some stationary or static state has been achieved.

\subsection{Ising model}

Also for human opinions, one could use the Ising model of the previous section
\cite{callen,galam}; see also \cite{weidlich}. People can vote for or against 
the government or a new constitution, for one of two presidential candidates, 
or (using generalised models) for one out of $Q>2$ different parties. Their 
neighbours on a lattice influence them in their vote, and in addition mass 
media may influence everybody in one direction. The latter effect can be 
modelled through an external ``magnetic'' field, eq.(1b) in ``Phase transitions
..." by this author in this encyclopedia. No motion of people needs to be
taken into account, and the complications of Kawasaki kinetics (exchange of 
two people with opposing opinions) are not needed. Thus the Glauber program
of the previous section still can be used, and we only refer here the old 
generalisation into the social impact model \cite{latane,schweitzer} and to a 
recent financial application \cite{sornette}.
 
\subsection{Voter model}

Also quite old is the voter model \cite{liggett}: Each person chooses between
two opinions, by taking over the one of a randomly selected neighbour. One
may rewrite this rule as stating that each person selects the opinion of the
neighbourhood, with a probability proportional to the number $n$ of neighbours 
having that opinion. Thus in contrast to the Ising model where the 
probabilities depend exponentially on $n$, now they depend linearly on $n$. 
A final equilibrium (absorbing fixed point) is reached if everybody shares the
same opinion. The deviations from that final state can be measured by the 
magnetisation (difference between the numbers for the two opinions) or energy
(average number of neighbours having the opposite opinion). The time needed to 
reach the consensus increases with a power of the lattice size, and the exponent
depends on the dimensionality. A nice and short review of the voter model, 
also on various networks, is given by the Majorca group \cite{sanmiguel}.

\subsection{Axelrod model}

Axelrod \cite{axelrod} wondered how different opinions or cultures may
coexists even if people tend to become more alike in their beliefs. Looking 
at the above Ising Figs. 1, 2, 3, we see that due to finite time and/or finite
temperature such coexistence of two opposing opinions is possible. But Axelrod
generalised it not only to $Q > 2$ different possible opinions as in the
Potts model of the ``Schelling'' section, but also to $F$ different questions.
People may have one set of opinions on which political party they want to vote
for, another set about what is the best football team, a third about recent 
cinema films, etc. This allows for $Q^F$ different opinion sets on all $F$ questions
(``features''). Of course, one could could generalise this model to the case 
where the number $Q_f$ of possible choices is different for the different
features $f$, allowing then for $\Pi_{f=1}^F Q_f$ instead of simply $Q^F$ 
different sets of opinions. 

Another aspect of the model takes into account that people prefer to talk to, 
or to make political coalitions with, others with whom they share many 
opinions. Thus the 
probability of one person to take over the different opinion of a neighbour
is proportional to the number of features on which their opinions already 
agree. In the next subsection we will use a similar concept under the name of 
bounded confidence.  

Whether a total consensus (``globalisation'') is reached or 
multiculturality persists depends on parameters: Small $Q$ lead to consensus.
Again, the Majorca group \cite{sanmiguel} reviewed the many follow-up papers
on this Axelrod model.

\subsection{Sznajd, Krause-Hegselmann and Deffuant models}

Much of the opinion dynamics research since 2000 centred on three different
models S, KH and D, originally invented independently around that year: 
Sznajd \cite{sznajd} (S), Krause-Hegselmann \cite{kh} (KH) and Deffuant et al 
\cite{deffuant} (D). They were also called missionaries, opportunists and 
negotiators by some computational physicists \cite{newbook}.

The S model is closest to the earlier models since it allows for $Q$
discrete opinions, while KH and D use real opinions, e.g. between zero and 
one. S happens on a lattice or network while for KH and D everybody may 
interact with everybody. In the most widespread S version a pair of neighbouring
sites on a square lattice convinces its six neighbours of its opinion, if
and only if the two opinions of the pair agree \cite{milgram}; governments and
parties usually lose support if their internal opinion differences make it to
the headlines. For KH, the new opinion of a person is the arithmetic average 
over the opinion of the whole population. For D, each person selects randomly 
another person and then both move in their opinion towards each other by an 
amount proportional to their opinion difference.

In all three cases, ``bounded confidence'' applies: The KH agents average only
over those people who differ from their own opinion by less than $\epsilon$, 
and the D agents only select negotiation partners differing by less than 
$\epsilon$ from their own opinions. In both models $0 < \epsilon < 1$
is a fixed parameter. For $S$ agents
with $Q=2$ such a rule makes no sense, but for $Q > 2$ one can modify the
convincing rule such that only neighbours differing by at most $\pm 1$ from
the pair opinion adopt the pair opinion. Thus $1/Q$ for S plays the role of
$\epsilon$ for D and KH. A rule similar to this bounded
confidence was mentioned above for the Axelrod model \cite{axelrod}.

Inspite of the differences in their definitions, the results are quite 
similar for S, KH and D. For large $\epsilon $ or $Q \le 3$ a complete 
consensus is usually reached; for small $\epsilon$ or $Q \ge 4$ different 
opinions may coexist forever. In addition to computer simulations, also 
analytical calculations were made \cite{redner,slanina} which agree with many
aspects of the simulations. More results, also for opinions on more than 
one feature and agents sitting on scale-free networks, are summarised by us in 
\cite{newbook}. 

One particular application is shown in Fig.5: Various Brazilian election results
for candidates in city councils showed great similarity if the number of 
candidates getting a given number $v$ of votes is plotted against this $v$. 
Putting the $S$ model with $Q = 1000$ candidates onto a scale-free network
instead of a square lattice, excellent agreement of simulation and reality
was found after the numbers were scaled by suitable factors. Also Indian
elections were simulated this way \cite{gonzalez}, while some exceptions
may also exist (S. Fortunato, unpublished). It would be nice to apply other 
opinion dynamics models to the same election problem. As usual in statistical 
physics, these studies can predict and simulate the shape of the distributions
but not the winner in a specific election, just as we can predict the pressure
of the air molecules around us but not where which molecule will be one minute
from now.

\begin{figure}[hbt]
\begin{center}
\includegraphics[angle=-90,scale=0.5]{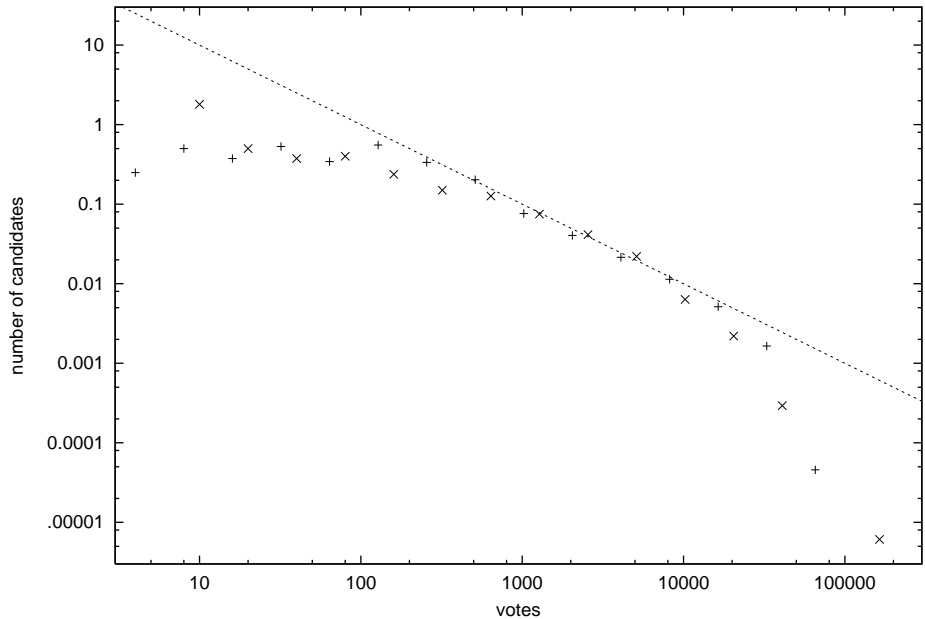}
\end{center}
\caption{Brazilian elections (x) and simulations of Sznajd model on Barab\'asi -
Albert networks (+); from \cite{bernardes}.
}
\end{figure}
\subsection {Galam conservatism}

Galam has published since many years theoretical models which may explain why
reforms are very difficult and why a minority can stay in power. Usually these
models are solvable analytically and assume that the population is divided into 
small groups of people which to the outside are represented by one person who
follows the majority wish of the group. Several of these representatives form
a supergroup, and this supergroup again decides according to the majority of
the representatives in it. In this way an ``infinite'' hierarchy of people,
groups, supergroups etc can be built. In the case of equally many voting for 
one choice as for the opposite choice, within one unit, that unit votes for the 
status quo. Starting with everybody having opinion --1, a very large majority
of people must switch to opinion +1 before the top of the hierarchy finally
also changes opinion \cite{galam2}. We refer to our book \cite{newbook} for 
a summary of more recent Galam papers.  
      
\section{Languages, Hierarchies and Football}

\subsection{Language competition}

Darwinian survival of the fittest is established biology, but similar concepts 
can be applied to human languages, bridging the gap to opinion dynamics. There
are now thousands of different languages, and their ``size'' is the number of
native speakers of that language. The size distribution extends from 1 (on the 
verge of extinction) to $10^9$ (Mandarin Chinese). The grammar of a language
\cite {wals} can be characterised by $F$ features each of which can have 
$Q$ different values, just as in Axelrod's model explained above. Features
can change spontaneously or be taken over from a (neighbouring) language; 
speakers of a small language give it up and learn a widespread language (as done
with physics research publications since 1945); people migrate to other places 
and bring their language with them. All these processes can lead to the 
extinction of existing languages and the creation of new ones (by the branching 
of one language into several daughter languages.) The present language size 
distribution is roughly log-normal, with an enhancement at small sizes
\cite{ethnologue}. Similar languages form families, and the size distribution
of families is a power law at intermediate sizes \cite{wichmann} (where the 
size is now the number of different languages) belonging to that family).

Various computer simulations of this language competition have been made, mostly
since 2003 and reviewed recently \cite{langssw}; see also
\cite{cangelosi,nowak}. We only mention Fig.6 from 
a modified Viviane model which agrees well with the real language size 
distribution. For language families, the empirical statistics is worse
\cite{wichmann} and the models work less well \cite{langssw}. 

\begin{figure}[hbt]
\begin{center}
\includegraphics[angle=-90,scale=0.5]{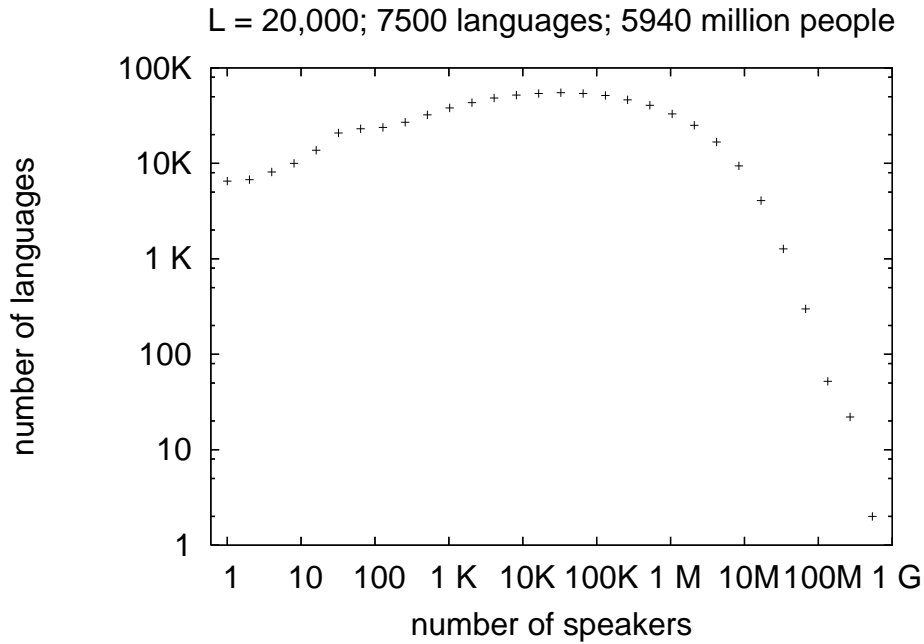}
\end{center}
\caption{Simulated language size distribution on a $20,000 \times 20,000$ square
lattice using a modified Viviane model \cite{pmco}.
}
\end{figure}

\subsection{Self-organisation of social hierarchies}

The elites of all countries and all times always had excellent reasons why 
they should be on top and others on the bottom. This holds even when the United 
Nations criticise the school system as violating human rights. In contrast, 
the Bonabeau model \cite{bonabeau} explains social hierarchies as purely 
accidental, without any merit. People are put on a lattice, occupying a 
fraction $p$ of all lattice sites and having an initial score of zero. Then 
they move randomly to neighbouring 
sites, and whenever one person wants to move into the site occupied by 
another person, a fight erupts. The winner takes the contested site, the loser 
moves into (or stays at) the other site. Also, the winner adds one point 
and the loser subtracts one point in its score, and in the future the agents
with a positive score have a higher probability to win, those with a negative
score have a lower probability to win. Slowly the history is forgotten, by 
reducing the score at each time step by, say, ten percent. 

With some suitable feedback between 
the distribution of scores and the probability to win, a phase transition was 
simulated such that for $p$ above some critical concentration, the standard
deviation in the scores becomes positive for long times and large populations.
For $p < p_c$ it fluctuates near zero, that means everybody remains near a 50 
percent chance to win. So, just by accident at a high population density
some people rise to the top, and others fall to the bottom. However, the people
on top (bottom) are not always the same; only the differences between top and 
bottom, not the people, remain the same. \cite{redner2,naumis,weisbuch} are
some of the more recent references in this field.
 
\subsection{Football}

Football (= soccer) is the world's most popular spectator sport, though in 
the author's city it is more a frustration. Randomness surely plays a role 
and makes it attractive. Can we explain all results just by chance, in the 
spirit of Bonabeau hierarchies? Assuming a constant probability to make a goal 
within one minute, the distribution of goals and victories is more narrow than 
in reality. If instead we assume that this probability varies from team to 
team, still no good agreement is found. Good agreement with reality is obtained
only if correlations are taken into account \cite{janke}, in the sense that 
a goal makes the scoring team happy, shocks the opposing team, and thus with an 
enhanced probability leads to another goal for the scoring team. Thus if we 
lose it is not just bad luck; it is also the referee's fault. 
 
\section{Future Directions}

The Schelling model of section 2 is not the only case of missed opportunities
because of a lack of cooperation between social sciences on the one side and 
physics, mathematics or computer science on the other side. The two books cited
at the end \cite{newbook} were written without the authors of one book knowing of
the preparation of the other book. One group of authors works in physics 
departments; none of the other group lists physics as institutional address. 
Nevertheless the two books show strong overlap in fields and methods covered,
but little overlap in the literature cited. More interdisciplinary cooperation
would help.

\end{document}